\documentclass[11pt,a4paper]{article}
\usepackage{moriond,epsfig}

\def\Journal#1#2#3#4{{#1} {\bf #2}, #3 (#4)}

\def\PLB{{\em Phys. Lett.}  B}
\def\PRL{\em Phys. Rev. Lett.}
\def\PRD{{\em Phys. Rev.} D}

\def\EPJ{{\em Eur. Phys. J.} C}

\def\be{\begin{equation}}
\def\ee{\end{equation}}
\def\bea{\begin{eqnarray}}
\def\eea{\end{eqnarray}}

\begin{document}
\vspace*{4cm}
\title{The electroweak fit and constraints on new physics}

\author{ Johannes Haller\footnote{for the {\it Gfitter} group (www.cern.ch/gfitter)} }

\address{Institut f\"ur Experimentalphysik,\\
Universit\"at Hamburg, Germany}

\maketitle\abstracts{ The global electroweak fit of the Standard Model
  (SM) with {\it Gfitter} can be used to constrain yet unknown SM
  parameters, such as the Higgs mass, but also physics beyond the SM
  (BSM) via the formalism of oblique parameters.  This paper presents
  updated results of the {\it Gfitter} SM fit using the latest
  available electroweak precision measurements and the recent
  combination of direct Higgs searches at the Tevatron. In addition,
  newly obtained constraints on BSM models, such as models with extra
  dimensions, little Higgs and a fourth fermion generation, are
  presented.  While a light Higgs mass is preferred by the
  fit in the SM, significantly larger Higgs masses are allowed in these new
  physics models.}

\section{Introduction}\label{sec:int}

By exploiting contributions from radiative corrections precise measurements
can be used to obtain insights into physics at much higher energy
scales than the masses of the particles directly involved in the
experimental reactions. In combination with accurate theoretical
prediction the experimental data allow us to constrain the free
parameters of the physics model in question.  Using this principle, in
particular the yet unknown mass of the Higgs boson $M_H$, can be
constrained in the Standard Model (SM) using the electroweak precision
measurements and state-of-the-art SM predictions since $M_H$ enters
logarithmically the prediction of higher-order corrections in the
SM. Furthermore, in models describing physics beyond the SM (BSM) new
effects, {\it e.g.}  from additional heavy particles entering the
loops, can influence the prediction of the radiative corrections of the
electroweak observables. The formalism of {\it oblique} parameters,
which parametrize the new physics contribution to the radiative
corrections, can then be used to probe the new physics models and
constrain their free parameters.

In this paper we present updated results of the global electroweak fit
with the {\it Gfitter} framework~\cite{gfitter} taking into account
the latest experimental precision measurements and the results of
direct Higgs searches from LEP and Tevatron. In addition, we present newly
obtained constraints on BSM models with extra
  dimensions, little Higgs and a fourth fermion generation 
  using the oblique parameters.

\section{The global electroweak fit of the SM with Gfitter}
\label{sec:ewfit}

A detailed discussion of the statistical methods, the experimental
data, the theoretical calculations and the results of the global
electroweak fit with {\it Gfitter} can be found in our reference
paper~\cite{gfitter}. Since its publication the fit has been
continuously maintained and kept in line with experimental and
theoretical progress. In the following the most important aspects of
the fit are quickly repeated and results of recent changes
-- mainly updates of the experimental data used in the fit,
{\it e.g.} $M_W$, $m_t$ and the direct Higgs searches at the Tevatron -- are
reported.

The SM predictions for the electroweak precision observables measured
by the LEP, SLC, and Tevatron experiments are fully implemented in
{\it Gfitter}. State-of-the-art calculations have been used, in
particular the full two-loop and the leading beyond-two-loop
corrections for the prediction of the $W$ mass and the effective weak
mixing angle~\cite{calc}, which exhibit the strongest constraints on
the Higgs mass. In the {\it Gfitter} SM library the fourth-order
(3NLO) perturbative calculation of the massless QCD Adler
function~\cite{n3lo} is included which allows to fit the strong
coupling constant with unique theoretical uncertainty.

\begin{figure}[t]
\begin{center}
\epsfig{file=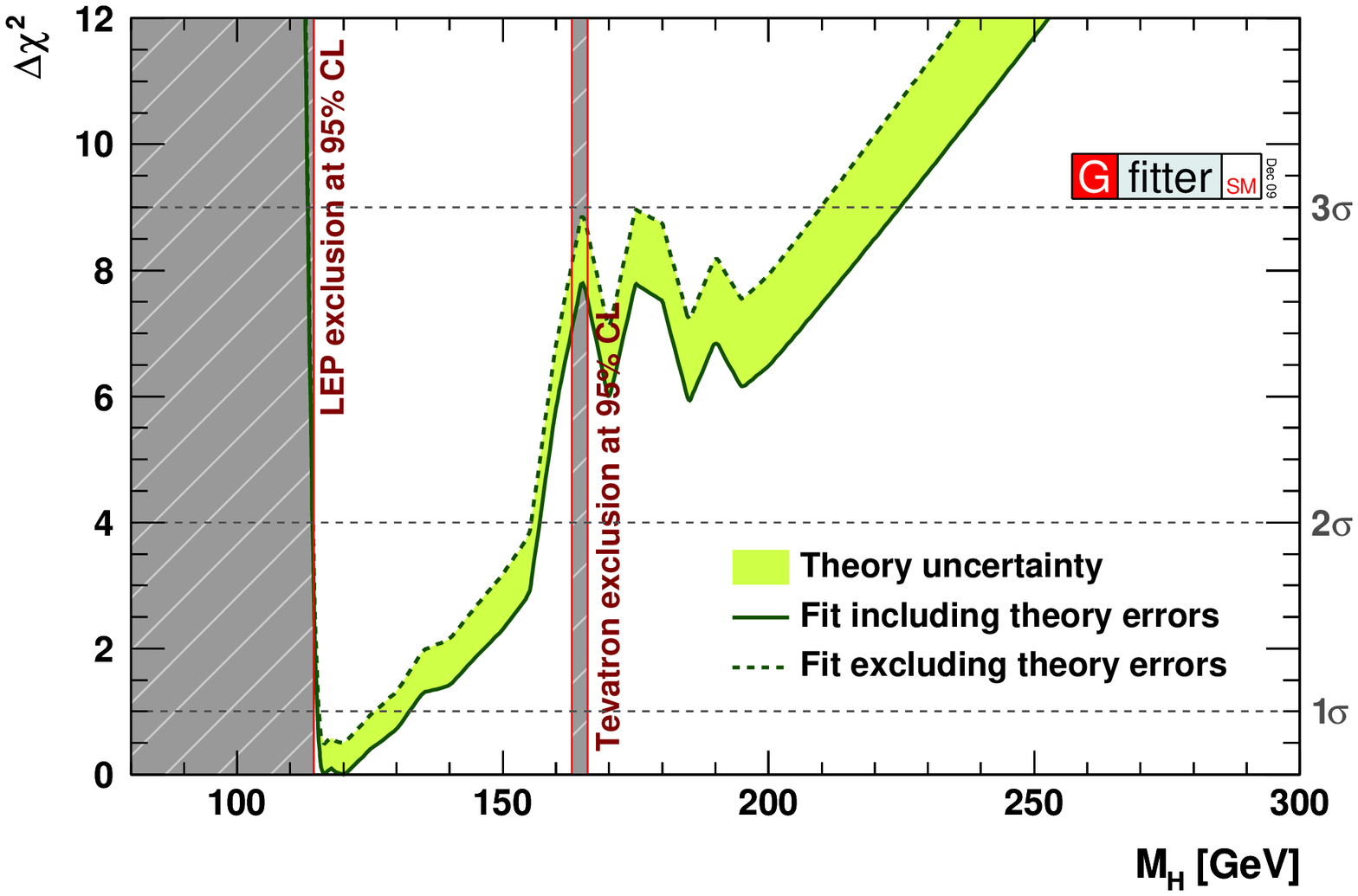,width=7.9cm}
\epsfig{file=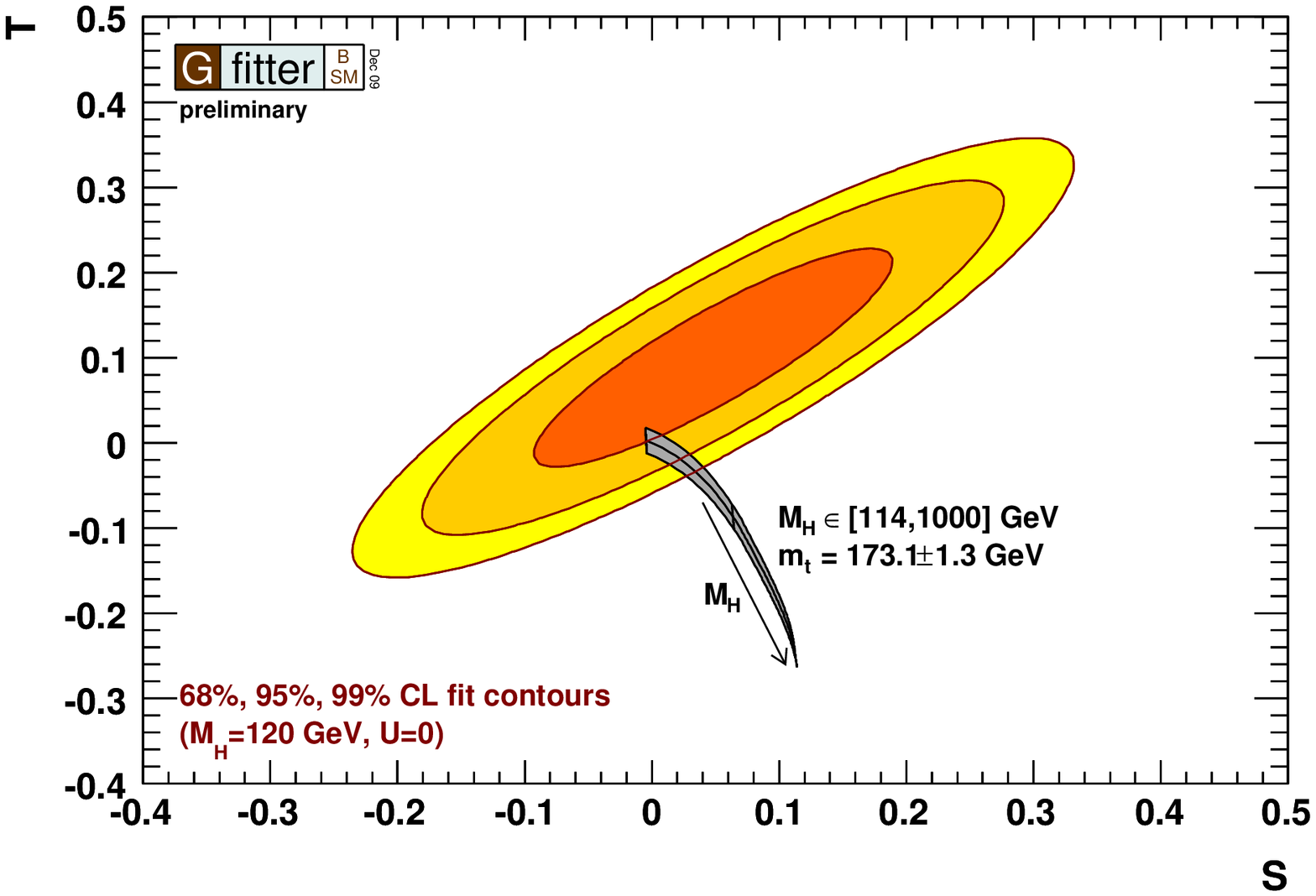,width=7.9cm}
\caption{(left) $\Delta\chi^2$ profile as a function of $M_H$ for the
  global fit of the electroweak SM with {\it Gfitter} including the
  results of the direct Higgs searches at LEP and Tevatron. The
  regions currently excluded with 95\,\% CL by LEP and Tevatron are
  indicated by the shaded areas.  (right) Fit result of the oblique
  parameters: Shown are the 68\,\%, 95\,\% and 99\,\% CL allowed
  regions in the $(S,T)$-plane with $U=0$ for a reference SM with
  $M_H=120$\,GeV and $m_t=173.2$\,GeV. The gray/dark area illustrates
  the SM prediction for various values of $M_H$ and
  $m_t$.}\label{fig:higgsscan}
\end{center}
\end{figure}

The experimental data used in the fit include the electroweak
precision data measured at the $Z$ pole~\cite{zpole}, the latest world
average~\cite{mw} of the $W$ mass $M_W=(80.399\pm0.023)$\,GeV, and
width $\Gamma_W=(2.098\pm0.048)$\,GeV, which include the recent run-2
mass measurement reported by D0, and the newest average~\cite{top} of
the Tevatron top mass measurements $m_t=(173.1\pm1.3)$\,GeV. For the
electromagnetic coupling strength at $M_Z$ we use the
$\Delta\alpha^{(5)}_{\rm had}$ value reported in ~\cite{dahad} which
does not include the recent ISR measurements of the $e^+e^-\to\pi^+
\pi^-$ cross-section from Babar and Kloe since an updated
$\Delta\alpha^{(5)}_{\rm had}$ value including both measurements is
not yet available. Also included in the fit is the information from
the direct Higgs searches at LEP~\cite{higgslep} and
Tevatron~\cite{higgstev}, where we use the latest combination.
\footnote{For the purpose of combination with the electroweak fit we
  transform the one-sided confidence level ${\rm CL}_{\rm s+b}$
  reported by the experiments into a two-sided confidence level ${\rm
    CL}^{\rm 2-sided}_{\rm s+b}$ and calculate the contribution to the
  $\chi^2$ estimator via $\delta\chi^2=2\cdot [{\rm Erf}^{-1}(1-{\rm
    CL}^{\rm 2-sided}_{\rm s+b})]^2$. A more detailed discussion of
  the combination method can be found in ~\cite{gfitter}. The
  alternative direct use of the test statistics $-2\ln Q$ in the fit
  leads to similar results.}

The free fit parameters are $M_Z$, $M_H$, $m_t$, $m_b$, $m_c$,
$\Delta\alpha^{(5)}_{\rm had}$ and $\alpha_S(M_Z^2)$ where only the
latter parameter is fully unconstrained since no direct experimental
measurement of $\alpha_S(M_Z^2)$ is used.  The minimum $\chi^2$ value
of the fit with (without) using the information from the direct Higgs
searches amounts to 17.8 (16.4) which corresponds to a $p$-value for
wrongly rejecting the SM of 0.22 (0.23). None of the pull values
exceeds 3$\sigma$. The 3NLO result of $\alpha_S(M_Z^2)$ obtained from
the fit is given by $\alpha_S(M_Z^2)=0.1193\pm0.0028\pm0.0001$, where
the first error is the experimental fit error and the second is due to
missing QCD orders. Among the most important outcomes of the fit is
the estimation of the mass of the Higgs boson. Without using the
information from the direct Higgs searches we obtain a $\chi^2$
minimum at $M_H=82.8^{+30.2}_{-23.3}$\,GeV with a 2$\sigma$ interval
of $[41,158]$\,GeV. The combination of the indirect fit with the
direct Higgs searches can be used to significantly reduce the allowed
regions for $M_H$ in the SM. The resulting $\Delta\chi^2$ profile as a
function of $M_H$ is shown in Fig.~\ref{fig:higgsscan}\,(left). The
expected strong increase at the LEP 95\% CL exclusion limit and the contribution of
the Tevatron searches at higher masses are clearly visible. We obtain
a $\chi^2$ minimum at $M_H=119.4^{+13.4}_{-4.0}$\,GeV with a 2$\sigma$
interval of $[114,157]$\,GeV.

\section{Constraints on new physics models}\label{sec:const}

A common approach to constrain physics beyond the SM using the global
electroweak fit is the formalism of oblique parameters. Assuming that
the contribution of new physics models only appears through vacuum
polarization most of the BSM effects on the electroweak precision
observables can be parametrized by three gauge boson self-energy
parameters ($S$, $T$, $U$) introduced by Peskin and
Takeuchi~\cite{STU}. In this approach the prediction of a certain
electroweak observable $O$ is given by the sum of the prediction of a
reference SM ($\rm SM_{ref}$, defined by fixing the values for $M_H$
and $m_t$) and the new physics effects parametrized by $STU$, {\it
  i.e.}  $O=O_{\rm SM, ref}(M_H, m_t) +c_SS+c_TT+c_UU$. The parameters
$STU$ hence measure deviations of the data from the chosen $\rm
SM_{ref}$. They vanish if the data are equal to the $\rm SM_{\rm ref}$
prediction. $S$ ($S+U$) is sensitive to BSM contributions to neutral
(charged) current processes at different energy scales, while $T$ is
sensitive to isospin violation effects. The parameter $U$ is small in
most BSM models. Further generalizations like additional corrections
to the $Zbb$ coupling~\cite{STUext} are also taken into account in
{\it Gfitter}.

Following this approach we have determined the oblique
parameters from the electroweak fit. For a $\rm SM_{\rm ref}$ with
$M_H=120$\,GeV and $m_t=173.2$\,GeV we obtain
\begin{equation} S=0.02\pm
0.11,\quad\quad\quad T=0.05\pm0.12; \quad\quad\quad U=0.07\pm 0.12\quad.
\label{eq:STU}
\end{equation}
The correlation between $S$ and $T$ is strong and positive ($+0.879$)
while the correlation between $S$ and $U$ and between $T$ and $U$ is
negative ($-0.469$ and $-0.716$,
respectively). Figure~\ref{fig:higgsscan}\,(right) shows the 68\%,
95\% and 99\% CL allowed contours in the $(S,T)$-plane for $U=0$,
together with the SM prediction featuring a logarithmic dependence on
$M_H$. Apart from the trivial fact that the prediction for our $\rm
SM_{ref}$ ($M_H=120$\,GeV, $m_t=173.2$\,GeV) is indeed $S=T=U=0$, it
can be seen that the data are compatible with the SM prediction for
small values of $M_H$. Hence, no actual need for new physics can be
derived from this study.

However, certain BSM models feature a similar agreement with the
data. The prediction of these models can cover large regions in the
$ST$-plane due to the allowed variation of the additional free model parameters
which in turn can be constrained by comparing the experimental data
and the model prediction. As
shown in the following, in some BSM models large values of $M_H$ are
allowed due to a possible compensation of BSM and Higgs effects.

\subsection{Universal Extra Dimensions}\label{ssec:ued}

\begin{figure}[t]
\begin{center}
\epsfig{file=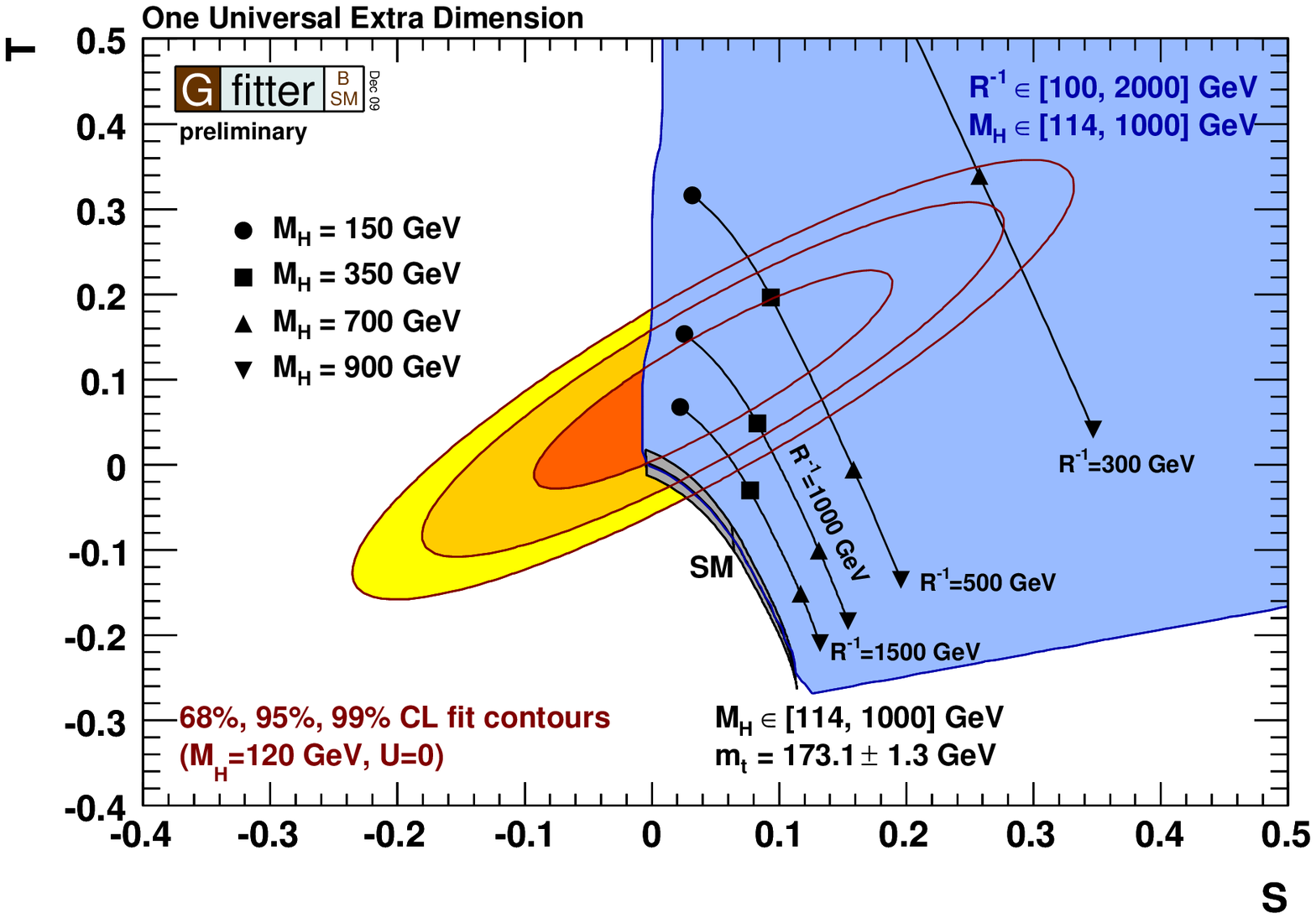,width=7.9cm}
\epsfig{file=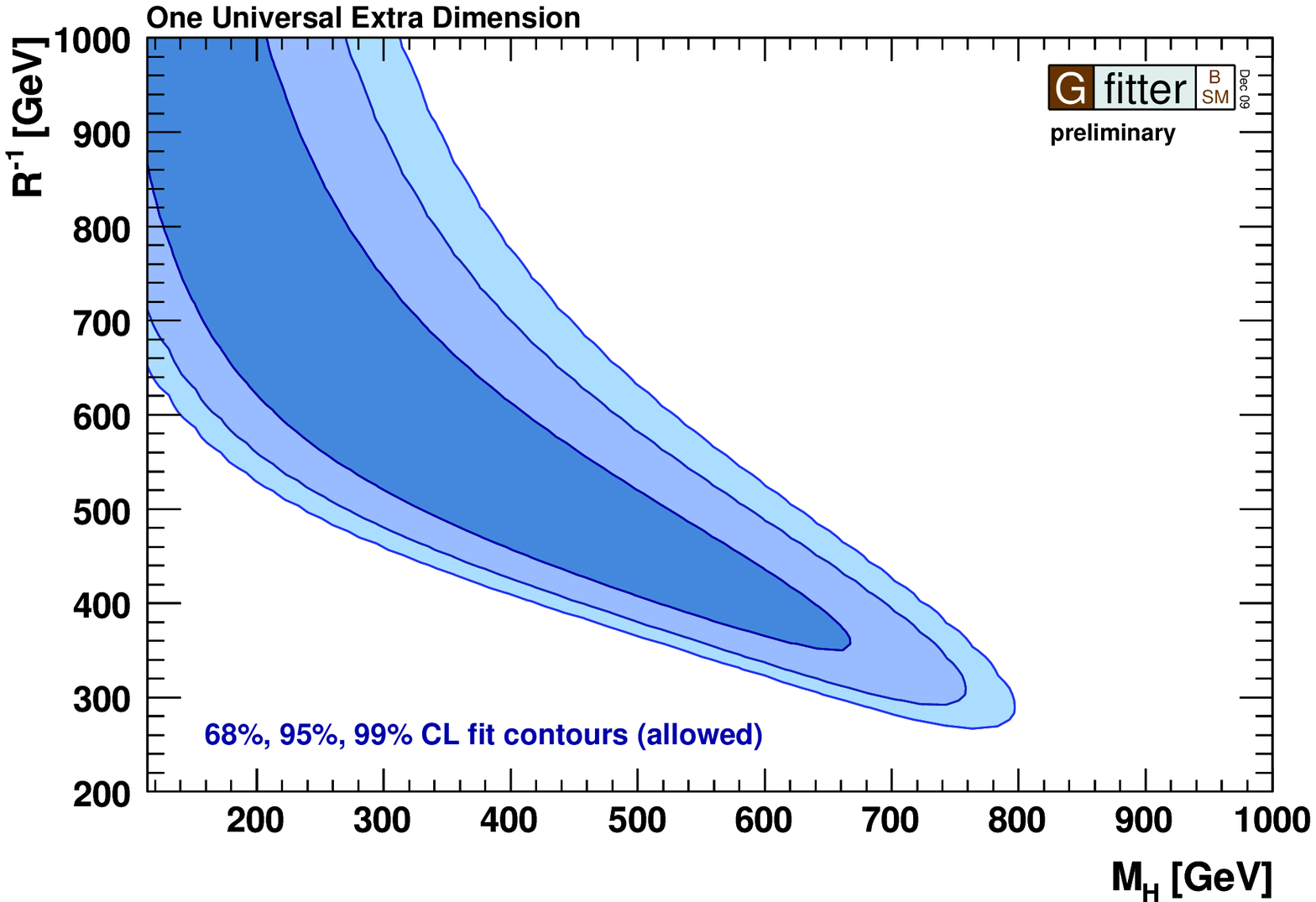,width=7.9cm}
\caption{Example results for a model with one Universal Extra
  Dimension: (left) Comparison of the $STU$-fit result with the
  prediction in the UED model for various values of the
  compactification scale $R^{-1}$ and the Higgs mass $M_H$. (right) The 68\,\%,
  95\,\% and 99\,\% CL allowed regions in the $(M_H,R^{-1})$-plane as
  derived from the fit.}\label{fig:stuued}
\end{center}
\end{figure}

As a first example we discuss a model with additional space dimensions
accessible for all SM particles~\cite{STUUED} (UED). In these models
the conservation of a Kaluza-Klein (KK) parity leads to a
phenomenology similar to supersymmetry with a stable lightest KK
state, which is a candidate particle for the cold dark matter in the
universe. The free parameters of the model are the number of extra
dimensions $d_{ED}$ and the compactification scale $R^{-1}$.  The
contribution to the electroweak precision observables via vacuum
polarisation effects in these models, {\it i.e.} the prediction of the
$STU$ parameters, have been calculated in \cite{STUUED}. The main
contribution results from additional KK-top/bottom and KK-Higgs
loops. For $d_{ED}=1$, as assumed in the following, the prediction of
the oblique parameters mainly depends on $R^{-1}$ and $M_H$.

In Fig.~\ref{fig:stuued}\,(left) the experimental fit result in the
$(S,T)$-plane is compared to the UED prediction for various values of
$R^{-1}$ and $M_H$. It can be seen that for high values of $R^{-1}$
the UED prediction approaches the SM expectation while for smaller
$R^{-1}$ values a significant deviation from the SM prediction is
expected. The same behavior can be observed in
Fig.~\ref{fig:stuued}\,(right) where the resulting 68\%, 95\% and 99\%
CL allowed regions in the $(M_H,R^{-1})$-plane are shown. For high
$R^{-1}$ values the constraint on $M_H$ approaches the SM result, {\it
i.e.} small $M_H$ are preferred, while for small $R^{-1}$ values,
significantly larger $M_H$ values are still allowed since the UED
contribution is compensated by a heavier Higgs boson. The latter parameter region 
is well within the direct discovery reach of the LHC since $R^{-1}$
indicates the expected mass region of the additional KK states.
The region $R^{-1}<300$\,GeV and
$M_H>750$\,GeV can be excluded at 95\% CL. These findings are in agreement with
previous publications~\cite{STUUED}.

\subsection{Littlest Higgs model with T-parity conservation}\label{ssec:lh}

Little Higgs theories tackle the SM hierarchy problem by
introducing a new global symmetry broken at a scale $f\sim1$\,TeV
where new SM-like fermions and bosons exist canceling the
one-loop quadratic divergengies of $M_H$ in the SM. The Littlest Higgs
(LH) Model~\cite{LH} is based on a non-linear 1$\sigma$ model
describing an SU(5)/SO(5) symmetry breaking. Similar to $R$-parity
conservation in supersymmetry, $T$-parity conservation provides a
possible cold dark matter candidate and,
important for the current discussion, it forbids tree-level
contribution from heavy gauge bosons to the electroweak
observables. In this case the dominant oblique corrections~\cite{STULH} rather
result from loops involving the two new heavy top states ($T$-even and
$T$-odd). The corrections depend on the scale $f$, the ratio of
the top state masses $s_{\lambda}=m_{T^-}/m_{T^+}$, $M_H$ and a coefficient $\delta_c$ whose exact value
depends on details of the UV physics. \footnote{The latter parameter is treated as theory
uncertainty in the {\it Gfitter} fit with $\delta_c=[-5,5]$.}

\begin{figure}[t]
\begin{center}
\epsfig{file=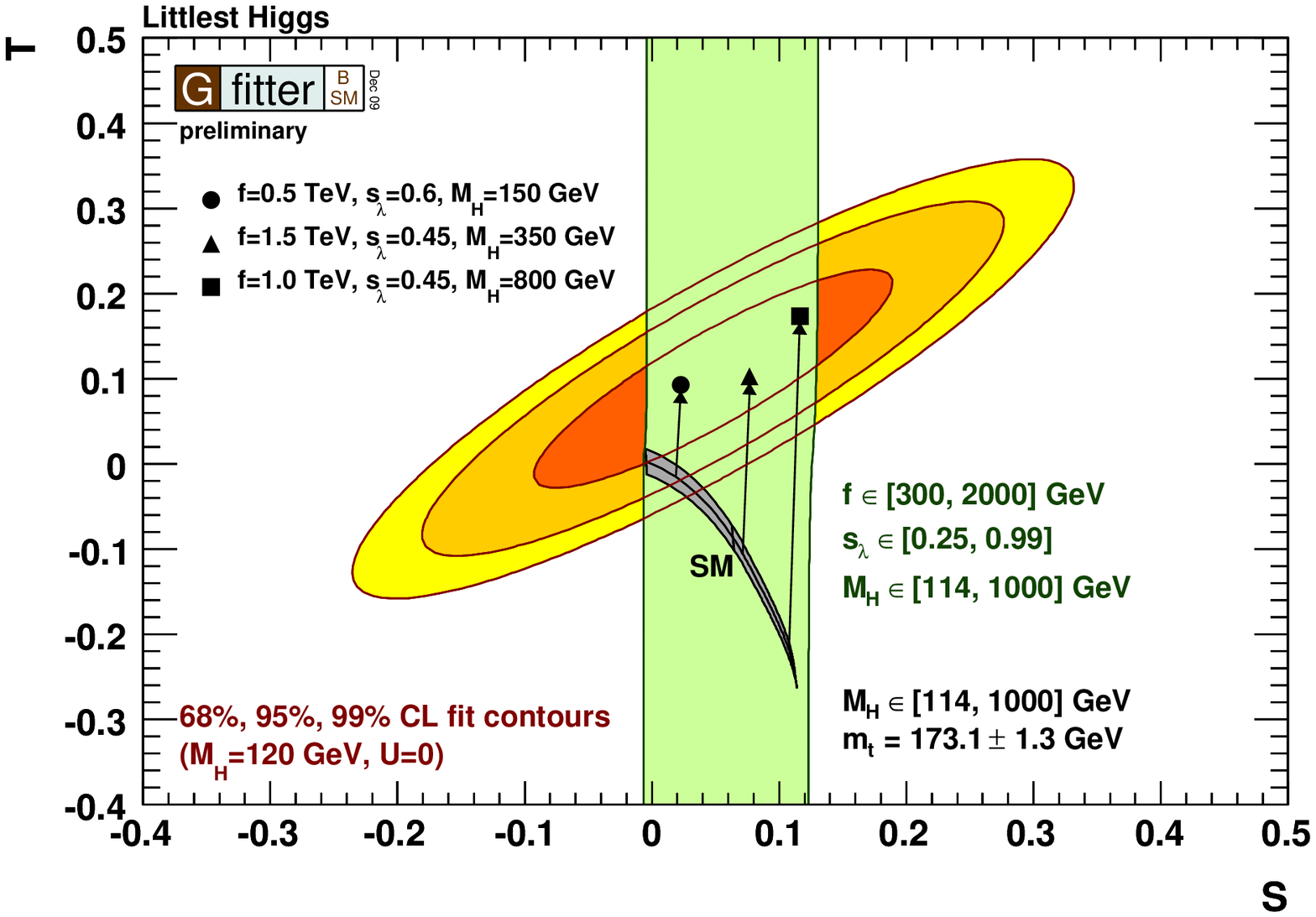,width=7.9cm}
\epsfig{file=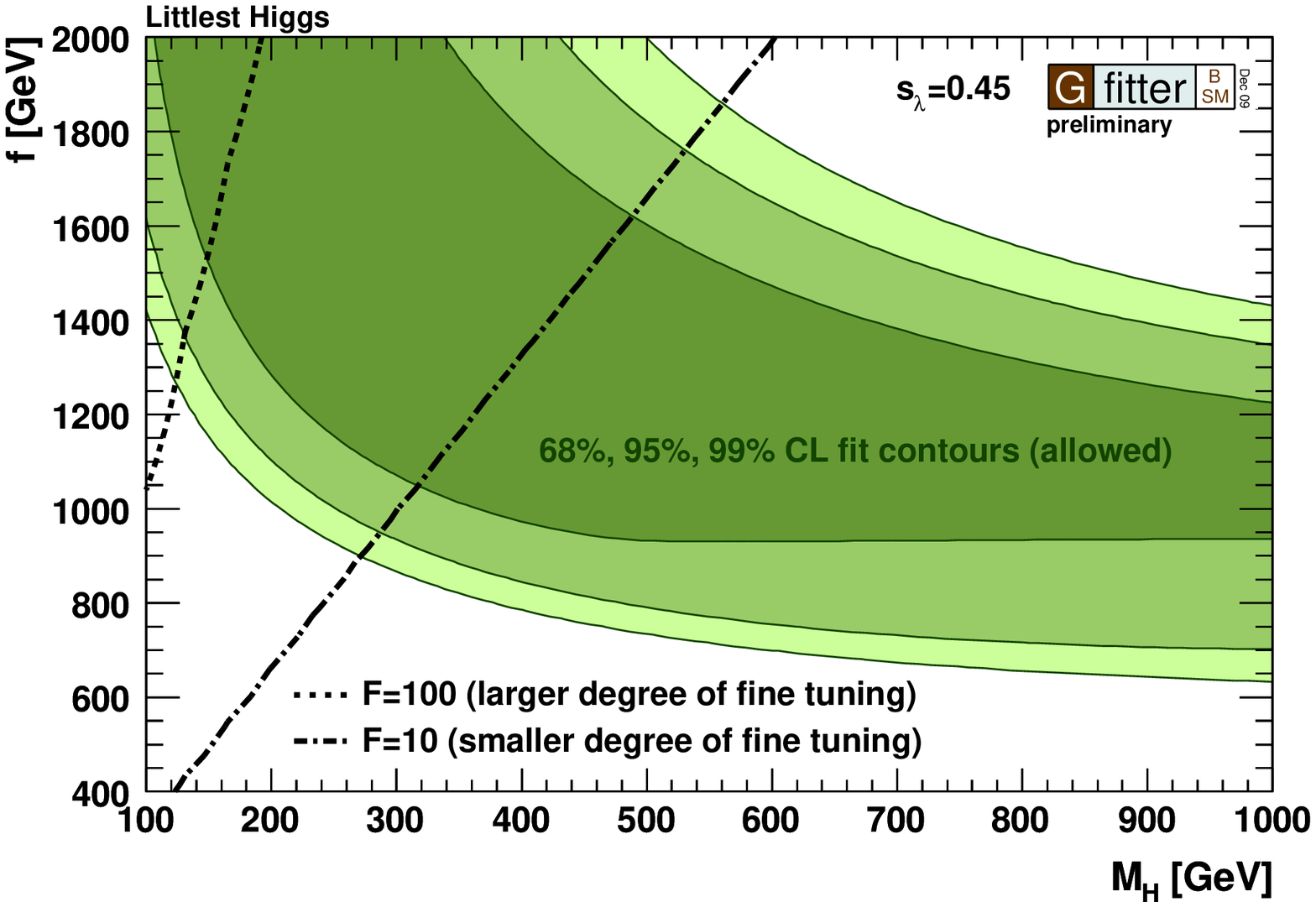,width=7.9cm}
\caption{Example results for the Littlest Higgs model with $T$-parity
  conservation: (left) Comparison of the $STU$-fit result with the
  prediction in the LH model. The symbols illustrate the LH
  predictions for three example settings of the parameters $f$,
  $s_{\lambda}$ and $M_H$. The light green area illustrates the
  predicted region when varying the free parameters in the ranges
  indicated in the figure.  (right) The allowed region in the
  $(M_H,f)$-plane as derived from the fit for
  $s_{\lambda}=0.45$.}\label{fig:stulh}
\end{center}
\end{figure}

In Fig.~\ref{fig:stulh}\,(left) the experimental fit result in the
$(S,T)$-plane is compared to the LH prediction for example values of
$f$, $s_{\lambda}$ and $M_H$. It can be seen that for certain
parameter settings the LH model with $T$-conservation is indeed in
agreement with the data.  In Fig.~\ref{fig:stulh}\,(right) the fit
results for $s_{\lambda}=0.45$ are illustrated as 68\%, 95\% and 99\%
CL allowed regions in the $(M_H,f)$-plane.  As expected, for high
values of $f$ the $M_H$-constraint in the LH model approaches the
$M_H$-constraint of the SM, while for smaller $f$ values significantly
larger values of $M_H$ are allowed than in the SM. Although the
allowed regions in the $(M_H,f)$-plane are strongly dependent on
$s_{\lambda}$ and no absolute exclusion limit on one of the parameters
alone can be derived, the above statements are true for all values of
$s_{\lambda}$.

\subsection{Models with a fourth fermion generation}\label{ssec:fg}

While the fermion sector of the Standard Model is composed of three
generations of leptons and quarks without explanation of this number,
several SM extensions suggest extra matter families.  In a simple,
generic model with only one extra family two new fermions
$(\Psi_1,\Psi_2)$ are added to both the quark and lepton sector, {\it
i.e.} a left-handed isospin doublet $\Psi_L=(\Psi_{1},\Psi_2)_L$ and
two right-handed isospin singlet states $\Psi_{1,R}$ and $\Psi_{2,R}$ with
charges equal to the three SM generations. The free model parameters are the
masses of the new quarks and leptons $m_{u_4}$, $m_{d_4}$,
$m_{e_4}$ and $m_{\nu_4}$ respectively.  Assuming no mixing of the
extra families among themselves and with the SM fermions the
additional one-loop fermionic contributions to the oblique corrections
have been calculated in~\cite{STU4th}. In particular, the
importance of an appropriate mass splitting of the up-type and
down-type fermions has been highlighted.

\begin{figure}[t]
\begin{center}
\epsfig{file=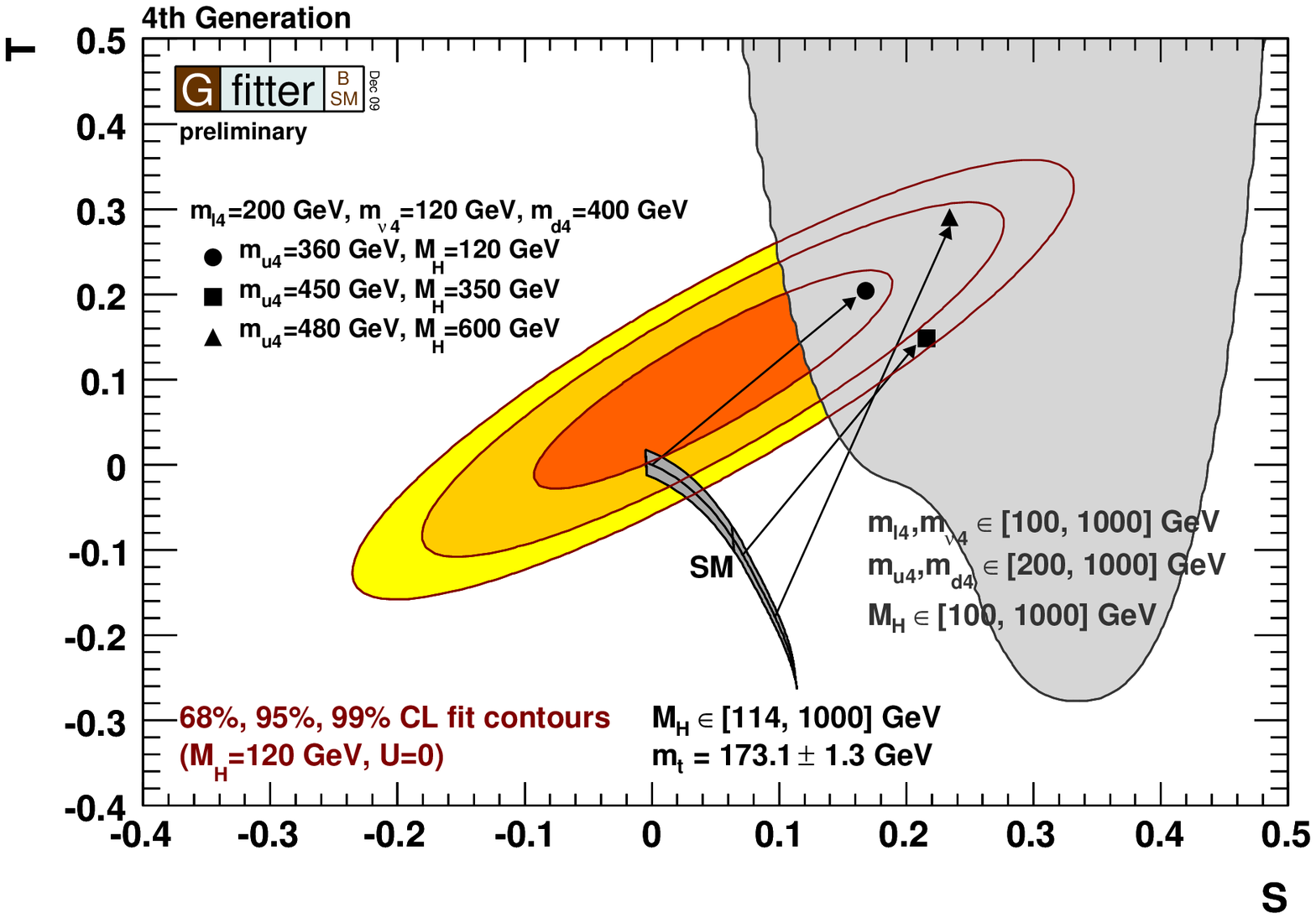,width=7.9cm}
\epsfig{file=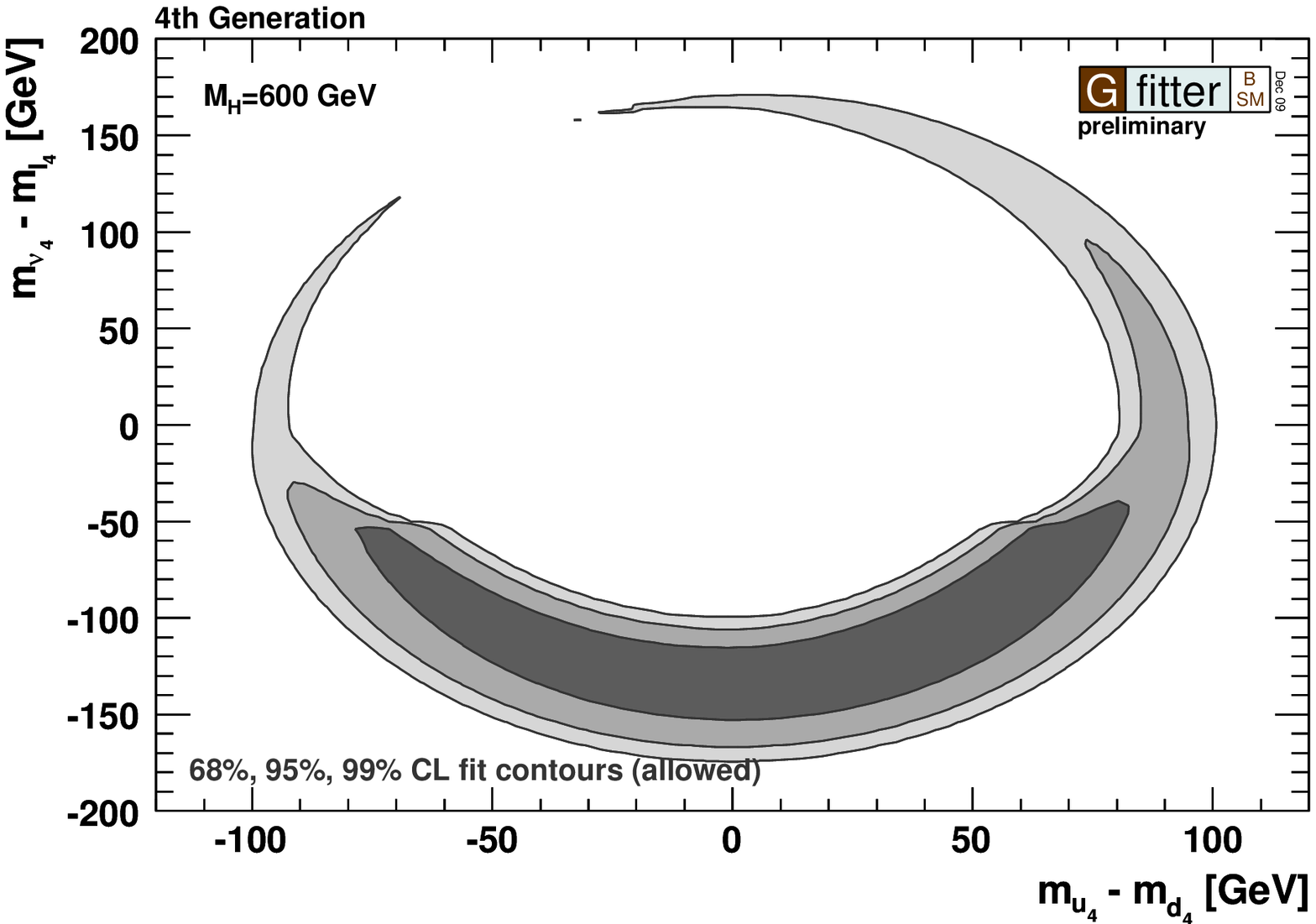,width=7.9cm}
\caption{Example results for a model with a fourth fermion
  generation: (left) Comparison of the $STU$-fit result with the prediction
  in the fourth generation model. The symbols illustrate the
  predictions for three example settings of the parameters $m_{U_4}$,
  $m_{d_4}$, $m_{{\nu}_4}$, $m_{l_4}$ and $M_H$. The light gray area
  illustrates the predicted region when varying the free parameters in
  the ranges indicated in the figure. (right) The allowed regions in
  the $(m_{u_4}-m_{d_4}, m_{l_4}-m_{{\nu}_4})$-plane as derived from the
  fit for $M_H=600$\,GeV.}\label{fig:stu4th}
\end{center}
\end{figure}

In Fig.~\ref{fig:stu4th}\,(left) our experimental fit result in the
$(S,T)$-plane is compared to the prediction of the fourth generation
model for example values of the masses of the additional fermions and
$M_H$. It can be seen that for some parameter settings the fourth
generation model is indeed in agreement with the data and high values
of $M_H$ could in principle be allowed. Since the oblique parameters
are mainly sensitive to the mass differences of the up-type and
down-type fermions and rather insensitive to the absolute mass values
of the additional fermions, we have calculated the 68\%, 95\% and 99\%
CL allowed regions in the $(m_{u_4}-m_{d_4},
m_{l_4}-m_{{\nu}_4})$-plane for various values of $M_H$. The example
results for $M_H=600$\,GeV, shown in Fig.~\ref{fig:stu4th}\,(right),
demonstrate that a high Higgs mass is indeed in agreement with the
data for a range of new fermion masses. In general, the data prefer a
heavier charged lepton.

\section{Conclusion and outlook}\label{sec:con}

Using the {\it Gfitter} package, the reimplementation of the global fit
to the electroweak precision data and its combination with the recent
results of the direct Higgs searches allows an exclusion of the SM
Higgs mass above 158\,GeV at 95\% CL. However, contributions from new
physics may change this result significantly. The effects on the
gauge boson self-energy graphs, called oblique corrections, are known
for most of the BSM models and must be continuously confronted with the
latest experimental data. Newly obtained results of a few example BSM models
implemented in {\it Gfitter} have been reported in this paper, demonstrating that larger $M_H$ values are in agreement with the electroweak precision data in these models.  Apart from an
continuous maintenance of the results reported here, an important
future objective of {\it Gfitter} will be a further diversification of
the latter analysis towards more BSM models.

\section*{Acknowledgments}
This work is funded by the German Research Foundation (DFG) in
the Collaborative Research Centre (SFB) 676 ``Particles, Strings and
the Early Universe'' located in Hamburg.

\end{document}